\documentclass[sigconf]{acmart}

\PassOptionsToPackage{table}{xcolor} %
\usepackage{hyphenat}
\usepackage{url}
\usepackage{xspace} %
\usepackage{amsmath}
\usepackage{mathtools}
\usepackage{amsfonts}
\usepackage{algorithm}
\usepackage[noend]{algpseudocode}
\usepackage{chngcntr} %
\usepackage[hide]{notes-alt}
\usepackage[table]{xcolor}
\counterwithout{algorithm}{section}
\usepackage{balance}
\usepackage{enumitem}
\usepackage{soul}
\usepackage{tikz}
\usepackage{tabularx}
\usetikzlibrary{fit,positioning,calc,shapes}
\usepackage{pgfplots}
\pgfplotsset{compat=1.13}
\usepgfplotslibrary{colorbrewer}
\usepackage{multirow}
\usepackage{breqn}
\usepackage{caption}
\usepackage{pifont}
\usepackage{stfloats}
\usepackage{booktabs}
\usepackage{subcaption}

\definecolor{cycle1}{RGB}{228, 26, 28}
\definecolor{cycle2}{RGB}{55, 126, 184}
\definecolor{cycle3}{RGB}{77, 175, 74}
\definecolor{cycle4}{RGB}{152, 78, 163}
\definecolor{cycle5}{RGB}{255, 127, 0}
\definecolor{cycle6}{RGB}{153, 153, 153}%
\definecolor{cycle7}{RGB}{166, 86, 40}
\definecolor{cycle8}{RGB}{247, 129, 191}

\newcommand{\cmark}{\textcolor{cycle3}{\ding{52}}} %
\newcommand{\xmark}{\textcolor{cycle1}{\ding{56}}}
\newcommand{\mehmark}{\mbox{\cmark\textsubscript{\kern-0.45em\tiny\xmark}}}

\newcommand{\mpara}[1]{\medskip\noindent{\bf #1}}

\newcommand{\problem}{\textsc{Maximum Preference Coverage}\xspace}

\newcommand{\bbranker}{$\mathcal{R}_{BB}$ }
\newcommand{\bbscore}[2]{$\mathcal{S}_{BB}(#1, #2)$ }
\newcommand{\explainer}{$\mathcal{R}_{E}$ }
\newcommand{\expscore}[2]{$\mathcal{S}_{E}(#1, #2)$ }

\newcommand{\random}{\textsc{random} }
\newcommand{\rankbiased}{\textsc{rank biased} }
\newcommand{\topkrankrandom}{\textsc{top-k + rank random} }
\newcommand{\topkrandom}{\textsc{top-k + random} }
\newcommand{\topk}{\textsc{top-k} }

\newcommand{\NP}{NP\xspace}

\newcommand{\fromto}{\longrightarrow}

\renewcommand{\vec}[1]{\boldsymbol{#1}}

\newcommand{\on}{\operatorname}
\newcommand{\sgn}{\on{sgn}}

\newcommand{\IR}{\mathbb{R}}

\newcommand{\cX}{\mathcal{X}}
\newcommand{\cD}{\mathcal{D}}

\DeclarePairedDelimiter{\indic}{\llbracket}{\rrbracket}
\DeclarePairedDelimiterX{\norm}[1]{\lVert}{\rVert}{#1}

\begin{document}
\title{Interpreting search result rankings through intent modeling}
\author{Jaspreet Singh}
\affiliation{%
  \institution{L3S Research Center}
  \streetaddress{Applelstraße 4}
  \city{Hannover.}   
}
\email{singh@l3s.de}

\author{Avishek Anand}
\affiliation{%
  \institution{L3S Research Center}
  \streetaddress{Applelstraße 4}
  \city{Hannover.}   
}
\email{anand@l3s.de}

\renewcommand{\shortauthors}{Singh and Anand}

\begin{abstract}

Given the recent interest in arguably accurate yet non-interpretable neural models, even with textual features, for document ranking we try to answer questions relating to how to interpret rankings. 
In this paper we take first steps towards a framework for the interpretability of retrieval models with the aim of answering 3 main questions ``What is the intent of the query according to the ranker?'', ``Why is a document ranked higher than another for the query?'' and ``Why is a document relevant to the query?''

Our framework is predicated on the assumption that text based retrieval model behavior can be estimated using query expansions in conjunction with a simpler retrieval model irrespective of the underlying ranker. We conducted experiments with the Clueweb test collection. We show how our approach performs for both simpler models with a closed form notation (which allows us to measure the accuracy of the interpretation) and neural ranking models. Our results indicate that we can indeed interpret more complex models with reasonable accuracy under certain simplifying assumptions. In a case study we also show our framework can be employed to interpret the results of the DRMM neural retrieval model in various scenarios.

\end{abstract}


\maketitle

\section{Introduction}
\label{sec:intro}



\mpara{Why Interpretability?} In the context of data-driven models, interpretability can be defined as ``\textit{the ability to explain or to present in understandable terms to a human}''~\cite{doshi2017towards}. For the most part, machine learnt models are used as black boxes which output a prediction, score or ranking without understanding completely or even partially how different features influence the output. In such cases when an algorithm prioritizes information to predict, classify or rank; algorithmic transparency becomes an important feature to keep tabs on to restrict discrimination and enhance explainability-based trust in the system. Recently, in the machine learning (ML) and NLP communities there has been growing interest in the interpretability of ML models~\cite{binder2016layer,nlp2015visualizing,lei2016rationalizing} but there has been limited work on interpreting retrieval models considered central to IR.  

\mpara{Learning complex ranking models.} When using machine learning models to rank results, the training data (clicks, human annotations) informs how signals/features should be combined. Various models have been employed, ranging from linear regression and decision trees to deep neural networks~\cite{mitra2017neuralCIKM,dehghani2017neural,hui2017position,desm16} more recently. On a more abstract level, by learning how to combine features to best rank documents, the machine learned model is indirectly encoding the query intent. Documents are then ordered by relevance, i.e. how well to do they match the underlying query intent.

In this paper we focus on interpreting complex ad-hoc ranking models like deep neural nets that automatically learn representations and features. These models are increasingly popular due to improved performance but their decisions are hard to interpret.

\mpara{Post-hoc Model Agnostic interpretability.} Approaches in interpretability can be broadly classified into models that are \emph{interpretable by design} and approaches that explain an already built model or \emph{post-hoc interpretability}. Models that are interpretable by design use training procedures and features that are themselves interpretable like decision trees, falling rule lists etc. However, they do not possess the modeling capacity, and hence are not as accurate as, of modern and more recent complex models. We therefore, focus on the post-hoc interpretability of learnt rankers that enables us to tackle a multitude of text based retrieval models irrespective of the underlying learning algorithm or training data.





\mpara{Interpreting intents in rankings. } We first ponder on what interpreting a rankings really means. Are we interested in explaining why a document was relevant or ranked above another ? Although that is certainly interesting, what we are really interested in is if a model is performing in accordance with the \emph{information need} of the user who issues the (usually underspecified) query -- a key concept in IR. In other words, what is the actual query intent as understood by the trained ranking model. 



There are two key benefits to uncovering the learned intent. First, users can immediately identify biases induced by varied training data. For example, consider the query \textit{afghan flag} and two different rankers:

\textbf{Ranker 1 Top Doc:} \texttt{American invasion leads to burning of afghan flags ...}

\textbf{Ranker 2 Top Doc:} \texttt{Americans share experience of visit to Afghanistan ...}

Both rankers seemingly prefer documents about Americans in Afghanistan but why they choose different top documents is unclear. However if we present the user with terms that encode the intent then this becomes easier to discern. The intent of \textbf{Ranker 1} is \{\texttt{war, america, conflict, soldier, defeat}\} and that of \textbf{Ranker 2} \{\texttt{pole, mast, hoist, tourist, nation, anthem}\}. Ranker 1 seems to be trained from a time period where the war in Afghanistan was more relevant. 

The second benefit is being able to help identify overfitting or under-performance. If spurious patterns like copyright messages are identified as part of the intent, developers can quickly rectify the data and improve the model. 

While terms are easy to understand as explanations, it may not be enough to accurately answer two important questions: Why is A relevant? Why is A ranked above B? Hence we additionally need an explanation model that can easily show how a document is relevant to the query using the intent terms. 
\todo{A line about how we model intents and what is our explanation. Details of the problem like maximization of pref. pairs can be left out}

\mpara{Our Approach.} A ranking task is different other supervised tasks in that it can viewed as an aggregation of multiple decisions -- either ordering multiple relevant documents or resolving pairwise preferences. This also makes applicability of previous works on simpler supervised tasks difficult.
We take a \emph{local} view to the problem of post-hoc interpretation of a black-box ranking model, i.e. interpreting the output for a query. Our approach is to learn a simple ranking model, with an expanded query, that approximates the original ranking. In doing so, we 
argue that the expanded query forms the intent or explanation of the query \emph{as perceived by the black-box ranker}. In coming up with the explanation or the intent we hypothesize that an expanded query along with a simpler and interpretable model, is an accurate interpretation of the black-box model if it produces a similar ranking to the ranking of the black-box model. Towards this we exploit the pairwise preferences between documents in the original ranking to extract relevant terms to be considered for expansion by postulating a combinatorial optimization problem called the \problem. 

We conducted experiments with the TREC Web Track test collection. We show how our approach performs for both simpler models with a closed form notation (which allows us to measure the accuracy of the interpretation) and neural models. Our results indicate that we can indeed interpret more complex models with reasonable accuracy under certain simplifying assumptions. In a case study we also show our framework can be employed to interpret the results of the DRMM neural retrieval model in various scenarios.
 



\section{Related Work}
\label{sec:related-work}

Interpretability in Machine Learning has been studied for a long time in classical machine learning. However, the success of Neural networks (NN) and other expressive yet complex ML models have only intensified the discussion.

On one hand this has largely improved performance, but on the other they tend to be opaque and less interpretable. Consequently, interpretability of these complex models has been studied in various other domains to better understand decisions made by the network -- image classification and captioning~\cite{captioningxu2015showattention,dabkowski2017real,imageclasssimonyan2013deep}, sequence to sequence modeling~\cite{seqalvarez2017causal,nlp2015visualizing}, recommender systems~\cite{interpretrecsysChang:2016} etc. 

Interpretable models can be categorized into two broad classes: \emph{model introspective} and \emph{model agnostic}. Model introspection refers to ``interpretable'' models, such as decision trees, rules~\cite{rulesletham2015interpretable}, additive models~\cite{caruana2015intelligibletrees} and attention-based networks~\cite{captioningxu2015showattention}. Instead of supporting models that are functionally black-boxes, such as an arbitrary neural network or random forests with thousands of trees, these approaches use models in which there is the possibility of meaningfully inspecting model components directly, e.g. a path in a decision tree, a single rule, or the weight of a specific feature in a linear model. 

Model agnostic approaches on the other hand extract post-hoc explanations by treating the original model as a black box either by learning from the output of the black box model, or perturbing the inputs, or both~\cite{ribeiro2016should,influencefunctionskoh2017understanding}. Model agnostic interpretability is of two types: local and global. \emph{Local interpretability} refers to the explanations used to describe a single decision of the model. There are also other notions of interpretability, and for a more comprehensive description of the approaches we point the readers to~\cite{lipton2016mythos}.



In information retrieval there has been limited work on interpreting rankings. Closest to our work is the recent work by Singh and Anand~\cite{singh2018posthoc} which tried to approximate an already trained learning to rank model by a subset of (the original features) interpretable features using secondary training data from the output of the original model. Firstly, it does not focus on interpreting intents and secondly it is not model agnostic limiting its usability.

\section{Explaining Rankings}
\label{sec:formal}





\textbf{Notation}: Assuming a collection of documents denoted by $\cD$, where
$\vec{d} \in \cD$ represents a document;
A \emph{ranking} problem given a query $q$ is specified by a finite relevant subset $Q = \{\vec{d}_1, \ldots, \vec{d}_k \} \subseteq \cD$ of documents, for some $k \in \mathbb{N}$, and the task itself consists of predicting a preferential ordering of these objects, that is, a \emph{ranking}. The latter is encoded in terms of a permutation $\pi \in \mathbb{S}_k$, where $\mathbb{S}_k$ denotes the set of all permutations of length $k$, i.e.,
all mappings $[k] \fromto [k]$ (symmetric group of order $k$).

A permutation $\pi$ represents the total order
\begin{equation}\label{eq:r}
  \vec{d}_{\pi^{-1}(1)} \succ \vec{d}_{\pi^{-1}(2)} \succ \ldots \succ
  \vec{d}_{\pi^{-1}(k)} \enspace ,
\end{equation}
where $\pi(k)$ is the position of the $k$th object $\vec{d}_k$, and $\pi^{-1}(k)$ the index of the document on position $k$ ($\pi$ is often called a \emph{ranking} and $\pi^{-1}$ an \emph{ordering}). We denote the top-k ranking for $q$ produced by a specific ranking model $R$ as $R(q) = \pi$.





\subsection{Posthoc Model Agnostic Interpretability}

We want to understand why for a query $q$ does the ranker \bbranker output the list \bbranker(q) as the top k documents from the matched documents retrieved from the index. The explanation $E_q$ then is a representation (visual/textual) of the underlying intent for $q$ which we hypothesize will help explain the behavior of \bbranker. In this work we are specifically interested in explaining adhoc rankers such as DRMM, DSSM and DESM that deal with only textual features but are hard to interpret primarily because of non-linear compositions of automatically learnt hidden features~\cite{mitra2017neuralCIKM,dehghani2017neural}. 


\textbf{Weak and Strong Model Agnosticism}: In this paper we specifically consider a post-hoc model agnostic setting to understand ranking decisions. In other words we operate on an already built retrieval model and we do not assume access to its learning algorithm and model parameters such as coefficients encoding feature weights or neural network parameters. This inherently renders the ranker \bbranker a blackbox that takes $q$ as input and outputs a permutation \bbranker(q)$=\pi$ over the retrieved documents. 

In fact, we consider two levels of model-agnostic interpretability -- \emph{weak agnosticism} and \emph{strong agnosticism}. In weak agnosticism, for a given query-document pair we can expect a relevance score from the underlying model. Since most rankers order documents in a pointwise manner this a natural assumption to make. Also, most approaches in the interpretability literature that try to explain black box models learned from a training procedure like document and image classification are in fact weakly agnostic. We also consider a stronger notion of model agnosticism that only expects a ranking from a model and does not assume any relevance scores. This scenario is also suited to models that output pairwise document preferences or a list directly.

In summary, for weak agnosticism we can expect to issue an arbitrary query $q$ and document $d$ as input to \bbranker and obtain the relevance score \bbscore{q}{d} whereas for strong agnosticism, we cannot get \bbscore{q}{d}.

\subsection{Explanation Model}

In this subsection we describe our explanation model $E_q$ consisting of two components --$T_q$ and \explainer.

\textbf{Explanation through terms ($T_q$)}: We consider terms as a natural way to explain the intent of the ranker for a given ranking decision. Terms in IR, words or phrases, are not only central to devising retrieval models but are also intuitive and understandable for humans. Hence we intend to construct an explanation model that generates a set of terms $T_q$ (unigrams in this work), akin to query expansions, for a given \bbranker given $q$. 

\textbf{Interpretable Ranker (\explainer)}: Retrieval models or rankers that induce a closed-form notation over well-understood IR features like term-level statistics, document lengths and proximity are in general more understandable and interpretable (like BM25, statistical language models~\cite{pontecroft1998language,lavrenko2017relevance}, DFR~\cite{amati2002probabilistic} etc). The explanation ranker or \explainer{} should be a ranker that computes the score of a document given a query independent of other documents in the list. \explainer should require only terms from a query and target document to compute the score. External feature values may confuse the user. Finally \explainer should compute the score of a document by considering simple linear operations on term frequency, term position, document length, term proximity and inverse document frequency values. We choose these features not only for the sake of simplicity but also because they have been shown to be essential indicators of textual relevance. Moreover such arguably simpler models are also amenable to query expansions\cite{lavrenko2017relevance,expansionsgong2005web}. 

We hypothesize that $E_q$ should be able to explain preferences in rankings of the form $\vec{d}_{\pi^{-1}(i)} \succ \vec{d}_{\pi^{-1}(j)}$ when $i>j$ with a simple, well-understood formulation of an explanation ranker \explainer in conjunction with $T_q$. Since \explainer is easy to interpret, we can answer questions such as why is document A ranked higher than document B by \bbranker by showing the user how the relevance scores are computed by \explainer given $T_q$.

\subsection{Fidelity of an Explanation Model} 
\label{sec:fidelity}
For an explanation to be convincing it should be faithful to the underlying model. Given a ranking \bbranker(q) to interpret or explain, the simpler explanation model $E_q$ should produce a prediction (ranking in this case) close to the target ranking. We refer to this closeness as \emph{fidelity}. We measure the fidelity of an explanation ranking by a rank correlation measure -- Kendall's $\tau$ as

\[
\tau(x,y,n)= \frac{2}{n(n-1)}\sum_{i\not=j} \sgn(x_i-x_j)\sgn(y_i-y_j)
\]

where $x \in $ \bbranker(q) and $y \in E_q$ are the black box and explanation rankings respectively, and $n$ is the size of each ranking. If $n$ is $k$ (top documents to explain) then we call this local fidelity. We theorize that an explanation is locally faithful to the underlying ranking model if it is able to capture the original intent of the ranker and thereby replicate its prediction. An $E_q$ with high fidelity should be able to explain a high number of document preference pairs. \emph{Local fidelity} refers to the ability of the $E_q$ to accurately explain preference pairs only from the top-k ranked documents while \emph{global fidelity} refers to all pairs from the retrieved document set.


\subsection{Problem Statement}

Formally, for a given keyword query $q$ and ranking \bbranker(q), produced by a complex black box ranker, \bbranker, we wish to identify a \emph{high fidelity local explanation} $E_q$. The fidelity of $E_q$ is measured by the rank correlation metric $\tau$ between \bbranker($q$) and \explainer ($q \cup T_q$). Essentially, for a selected \explainer, $T_q$ is the set of terms that best preserves $\pi$ given $q$ and \bbranker.



\todo{introduce global fidelity. Connect the formalism better to the text and probably give formula for global and local fidelity in terms of $\tau$.}

\todo{Need to enumerate our assumptions relating to importance of terms to documents and relative importance of documents that we operationalize later}

\section{Approach}
\label{sec:pref-opt}


To explain the top-k documents we should be in principle be able to explain all preference pairs $\vec{d}_{\pi^{-1}(i)} \succ \vec{d}_{\pi^{-1}(j)}$ where $i < j \leq k$. In what follows we try to identify terms $\vec{w} \in T_q$ such that an expanded query $q \cup T_q$ using \explainer preserves most of the preference pairs.

Towards this, we make another simplifying assumption based on the importance of a term given a document and query that typically holds true in many reasonably well performing classical retrieval models. We assume that if a term $\vec{w}$ induces $\vec{d}_{\pi^{-1}(i)} \succ \vec{d}_{\pi^{-1}(j)}$ then  $\vec{w}$ is more relevant to $\vec{d}_{\pi^{-1}(i)}$ than $\vec{d}_{\pi^{-1}(j)}$ given the query assuming all terms are independent of each other. This indeed holds true for many classical term-based retrieval models like language models and BM25 where terms are assumed to be independent of each other and the overall relevance score is computed as an aggregation of \emph{only relevant} terms. In fact, we use one of these simpler models as our \explainer where this assumption strictly holds.


\textbf{Optimizing preference pair coverage}: We start with a reference set of expansion terms denoted by $\cX$, where each expansion term $\vec{w} \in \cX$ is described by a feature vector; thus, a term is a vector $\vec{w} = (p_1, \dotsc, p_d) \in \IR^d$, and $\cX \subseteq \IR^d$. Each feature corresponds to a preference pair $\vec{d}_{\pi^{-1}(i)} \succ \vec{d}_{\pi^{-1}(j)}$ and its value determines to what degree is the preference pair satisfied if $\vec{w}$ is chosen (described in detail in Section~\ref{sec:scoring}). We build the preference matrix $P$ from the term vectors and intend to find a minimal set of terms $T_q \subseteq \cX$ or simply $T$ as expansion terms. 

\begin{definition}[Preference Coverage]
  Given a selection set represented as a boolean vector $\vec{s}$, the \emph{preference coverage}  \textsc{PCov} over the aggregate vector $\vec{y} = \vec{s}^T\mathbf{X}$ is given by 
$\textsc{PCov} (\vec{s}) = \sum_i{\indic{y_i > 0}}$.
  
  \end{definition}

\begin{figure}[ht]
\centering
	\includegraphics[width=0.3\textwidth]{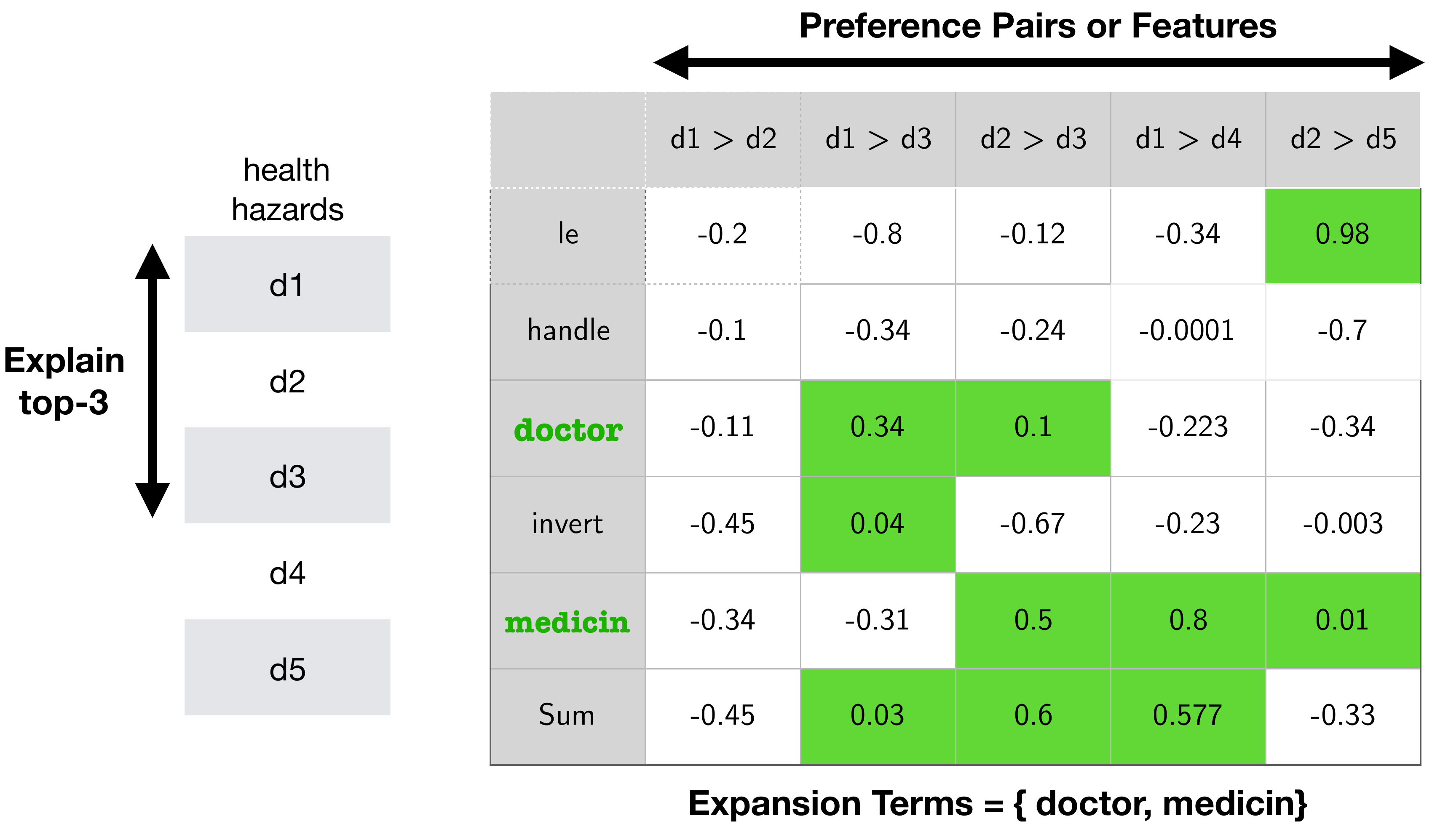}
    \caption{preference Matrix for the ranking for the query \textit{health hazards} }.
    \label{fig:prefmat}
\end{figure}

\todo{The diagram has a wrong cell -0.0001. Annotate the features and terms. }

The best selection of expansion terms naturally is the set that maximizes the preference coverage or explanation fidelity. We pose the \problem problem as choice of $S_{max}$ such that maximum preferences are covered. Writing it as an Integer Linear Program we have:
\begin{align}
\max &  \sum_j{\indic{y_j > 0}} \\
	       s.t.  & \\
 		   & x_i \in \{ 0,1\} \\
           & y_j = \sum_{0 \leq i \leq n} x_i.P_{i,j}
\label{eq:ilp}
\end{align}




\begin{proposition}
The \problem problem is \NP hard.
\end{proposition}
We do not include the proof in the paper for space reasons but we note that the proof sketch follows from the fact that the \problem is a generalization of the well known \textsc{Set Cover} problem. Not only is the \problem problem \NP hard, it is also easy to see that it is not sub-modular. This is clearer from the example in Figure~\ref{fig:prefmat}. Say the term \texttt{medicine} is chosen into the selection set with a \textsc{PCov}(\{\texttt{medicine}\}) = 3. Now chosing the term \texttt{handle} infact reduces \textsc{PCov}(\{ \texttt{medicine}, \texttt{handle} \}) = 2.


\subsection{Greedy Algorithm}
\label{sec:greedy}

Although \problem does not induce submodularity, we propose a heuristic greedy algorithm that intends to maximize the preference coverage of the input. At each iteration the algorithm greedily choses the term into the selection set that provides the maximum utility. The utility of a term $\vec{w}$ at any stage of the algorithm $U(\vec{w}, T)$ is the increase in the preference coverage when $\vec{w}$ is added to the selection set $T$ or 
\[
	U(\vec{w}, T) =  \textsc{PCov}(T \cup \vec{w}) - \textsc{PCov}(T) 
\]

In case of ties, the $\vec{x}$ that has the highest column sum (denoted by $Psum$) of the covered features are considered. 

\[
	Psum(\vec{w}) =  \sum_{w_i > 0 \& w_i \in \vec{w}} w_i 
\]


            


Now we turn to two important modeling choices that determine the input to the \problem: 

\begin{enumerate}
\item \emph{How do we select candidate terms for the expansions ?}
\item \emph{Which preference pairs do we model as features ?} and 
\item \emph{What should be the feature value or strength of a preference pair given a candidate term ?}.
\end{enumerate}

\subsection{Modeling Preference Pairs}
\label{sec:pref-pairs}

The \problem with concordant document pairs as features naturally tries to maximize fidelity as defined by Kendall's-$\tau$. Instinctively, the number of preference pairs to be considered in explaining top-k documents would be $k.(k-1)/2$. However, only considering the preference pairs from the top-k documents could result in higher likelihood of false positive terms being included in the selection set. \emph{False positive terms} tend to be general terms that co-occur with relevant terms resulting in an increased local fidelity (local here refers to the top ranked documents). We counteract the effect of false positives by sampling additional preference pairs from outside the top-k and requiring that the algorithm covers a larger number of preference pairs. In other words, high local fidelity alone acts akin to overfitting and additional preference pairs tries to achieve global fidelity by actively trying to reduce false positives added to the selection set.
We also hypothesize that carefully sampling pairs can help focus on other important query aspects and prevent overfitting. Conversely, improper sampling or sampling too many pairs from the tail could lead to good global rank fidelity but poor local rank fidelity. 


\mpara{Sampling Preference Pairs/Features.} As discussed in Section~\ref{sec: pref-pairs} considering all preference pairs is computationally prohibitive. Instead we sample pairs by the following strategies :

\begin{enumerate}
\item \random: Randomly select preference pairs from the target ranking.
\item \rankbiased: Sample preference pairs from the target ranking that are weighted by rank. Each pair is weighted by $1/rank(d_i) + 1/rank(d_j)$.
\item \topkrankrandom: Construct preference pairs based on a combination of rank and random sampling. In this method, for a preference pair $(d_i > d_j)$, $d_i$ is first rank bias sampled and then $d_j$ is randomly sampled. In addition, we consider all pairs from the top-k results to explain.
\item \topkrandom:  We consider all pairs from the top-k results to explain and a fixed number of randomly sampled pairs.
\end{enumerate}
 We contrast these against \topk (no sampling; consider only all pairs from the top-k results to explain) in the experiments.

\todo{Explain the exhaustive preference pair generation at top-k ranks. Explain clearly the effect of overfitting with examples and the need for sampling from the tail. Introduce a few sampling strategies. JS: wont the examples come in the experiments?}

\subsection{Preference Pair Scores}
\label{sec:scoring}


For each preference pair / feature, we compute a score that encodes the degree of concordance the candidate term maintains for the pair. We employ \explainer to first estimate the importance of a term for each document. The score for the term $\vec{w}$ given a pair $\vec{d}_{\pi^{-1}(i)} \succ \vec{d}_{\pi^{-1}(j)}$ is computed as 
\expscore{\vec{w}}{\vec{d}_{\pi^{-1}(i)}} - \expscore{\vec{w}}{\vec{d}_{\pi^{-1}(j)}}



This allows us to directly optimize for fidelity if we select an \explainer akin to a language model where $q_i \in q$ are terms in a query.
\begin{equation}
P(d|q) = \prod_{q_i \in q} P(q_i|d) = \sum_{q_i \in q} - log P(q_i|d)
\end{equation}
Now selecting terms that maximize coverage of concordant pairs is equivalent to selecting terms that when used as expansions closely reproduces \bbranker(q). A positive score in for a feature indicates that a term is more relevant for $\vec{d}_{\pi^{-1}(i)}$ than $\vec{d}_{\pi^{-1}(j)}$ according to R where $\vec{d}_{\pi^{-1}(i)} \succ \vec{d}_{\pi^{-1}(j)}$. For a set of terms, simply adding the scores for that feature will indicate if concordance is maintained and in turn maximize fidelity.





\subsection{Candidate Term Selection}
\label{sec:cand-selection}

In the previous section we presented the scheme for selection of the explanation terms. However, constructing a preference matrix over the entire vocabulary or a large fraction is neither desirable nor practical. In this section we propose approaches to meaningfully chose a set of candidate terms that should be considered as candidates for the \problem. 

\mpara{Initial candidate selection} For both, weakly and strongly agnostic cases, we first select a set of terms $C_{I}$ from the retrieved documents. We hypothesize that the frequently occurring terms that are not generic (according to the collection -- like stop words) are crucial to the rank order induced by \bbranker. Each term in the initial retrieved set is first scored using TF-IDF. TF is the frequency of term occurrence in the retrieved set and IDF can be estimated on an external corpus. Then the top-k highest scoring terms are selected (we set k to 1000 in our experiments). 

Not all chosen terms will contribute to the explanation and many of them would be false positives. We may encounter terms that co-occur with the true explanation terms which are hard to distinguish due to the reliance on \explainer to compute the term score for a given pair. To identify and remove these difficult false positive terms we resort to systematically modifying the document to isolate false positives. We call this document perturbation and in what follows we discuss two strategies: \emph{reductive} and \emph{additive} perturbation based on the hypothesis that removal and addition of non-relevant terms does not change the relevance of the document to the query by a large extent. Note that document perturbation can only be applied in a weakly agnostic setting.

\mpara{Reductive Perturbation} Given $C_{I}$, we perturb each document $d$ from a sample of documents from by removing a candidate $\vec{w} \in C_{I} \wedge \vec{w} \in d$. Due to retrieval models being sensitive to document length in general, we do not plainly remove all occurrences of $\vec{x}$ but instead replace each removed word with an out of vocabulary term. This allows us to mitigate the effect of length normalization. The perturbed document $d'$ is then scored by using the blackbox ranker to output \bbscore{q}{d'}. We then select the top $\vec{w} \in C_{I}$ that on average decrease the score of the document after reductive perturbation, i.e., (\bbscore{q}{d} - \bbscore{q}{d'}) $> 0$. 





\mpara{Additive Perturbation} Reductive perturbation is inherently a local approach and only deals with the terms present in the document. To further reduce the false positives, we use additive perturbation where we systematically add terms (absent from the document) from the vocabulary or initial candidate list $C_I$. For each $d$, we add $\vec{w}$ $n$ times to a document. $n$ can be a fixed constant or decided dynamically for each document. Akin to reductive perturbation, $S_{BB}(q, d')$ is computed for all perturbed documents. The top terms that cause the highest score increase , i.e. (\bbscore{q}{d'} - \bbscore{q}{d}) $ > 0$) on average are retained as candidates. Additive perturbation allows us to control not only the frequency of the added candidate term but also the position and order. 

In general (from our experiments) we observed that additive perturbation is more expensive to compute than reductive perturbation since we typically have more terms to add from the candidate list than to remove from a document. In our experiments we use a combination of the two to efficiently select a set of candidates with minimal potential false positives.


Both the perturbation approaches try to trade-off precision for recall by actively trying to control for false positives. It is also worthwhile to note that each of the perturbed documents could be considered as being present in the local neighborhood of the original document. Our hope here is that the \bbranker does similar predictions in the same neighborhood and thereby we try to exploit the locality of decisions.  

Once we construct a good set of candidates we build the preference matrix corresponding to a preselected \explainer. In this paper we focus on finding $T_q$ given a single preference matrix in order to thoroughly evaluate our proposed approach. However we can also construct multiple preference matrices corresponding to different \explainer (based on smoothing, position, proximity, etc.) and then find $T_q$. We leave this direction open for future work.




\section{Experimental Setup}
\label{sec:setup}

In order to establish the effectiveness of our approach, we answer the following research questions in our experimental evaluation:

\begin{itemize}

\item \textbf{RQ I:} What is the quality of our explanation approach in terms of fidelity and accuracy ?


\item \textbf{RQ II:} Which is the best sampling strategy for feature generation ? 


\item \textbf{RQ III:} Is our approach useful in identifying biases and  explaining document pairs? 
\end{itemize}

\mpara{Challenges in Evaluation.} 
The major challenge obtaining ground truth for \emph{perceived intent of a black-box ranker} (not of the actual user or query intent) is difficult if not outright impossible. In order to quantify the quality of our explanations we resort to black-boxes whose intents are fully understood. This infact is common practice in evaluation of local posthoc-interpretable approaches~\cite{ribeiro2016model,kim2016examples} with the underlying assumption that if an explanation model can correctly locally interpret a simple well understood model then it can faithfully (locally) interpret other complex models.

\mpara{Dataset and Queries.} For all experiments we use the web track Clueweb09 TREC test collection (category B) for adhoc retrieval. We use all 300 queries for the strongly agnostic case and the first 200 queries for the weakly agnostic case. Note that in our experiments we are more interested in showing that our approach can be applied to a variety of retrieval models (trained on a large training set) rather than the same retrieval model across multiple test collections. For that reason we consider only one Web scale collection with a large set of queries rather than more number of datasets.


\mpara{Metrics.} We use the following metrics to measure the efficacy of our approach.
\begin{itemize}
\item \textbf{Fidelity} (as introduced in Section~\ref{sec:fidelity}) between \bbranker(q) and \explainer($q \cup T_q$) measured by Kendall's $\tau$. 
For local fidelity we use $\tau$@10 and for global fidelity we use $\tau$@1000. 
\item \textbf{Accuracy} of the explanation by computing the fraction of $T_q$ terms that overlap with the set of ground truth terms $\mathcal{G}_q$. 
\item \textbf{Recall} computed as the fraction of candidate terms (from which $T_q$ is selected) that overlap with $\mathcal{G}_q$. 


\end{itemize}

We select the top-10 documents to explain for a given $q$ and \bbranker and set max $|T_q|$ to 10 in all our experiments.



\mpara{Blackboxes Ranking Models.} As described in our evaluation rationale we use a set of (simpler) retrieval models that whose intents are already known. Specifically, we chose the retrieval models that use pseudo-relevance feedback for re-ranking top-k documents and those that use word embeddings for expansions. Since both these retrieval models use some kind of expansions for a second round of scoring we assume these expansion terms as ground truth for our explanations. We chose exactly 10 expansion terms per query as ground truth for each ranker.

We use 4 blackbox models in our experiments. We first devised 2 different (non-neural) black box rankers -- RM3 and EMB. For each we use a language model with Jelinik-Mercer smoothing with $\alpha=0.4$ as $\mathcal{R}$. We specifically chose this setting to make it different from \explainer (described later).

\begin{equation}
P_{\mathcal{R}}(d|q) = \prod_{q_i \in q} \alpha P(q_i|d) + (1-\alpha)P(q_i|\cD)
\end{equation}

For the expansions we used two distinct approaches: pseudo-relevance feedback and word embeddings. We chose exactly 10 expansion terms per query for each of the following black-boxes: 

\textbf{RM3} Using the RM3~\cite{lavrenko2017relevance} algorithm, we determined a set of relevant expansion terms from the top-k documents for a given query. We then re-ranked the results using the aforementioned language model. We use RM3-k where $k = \{10,20\}$.

\textbf{EMB}  Instead of using pseudo-relevant documents to find expansion terms, an external collection is used for the expansions. We use glove embeddings (300 dimensions) trained on English Wikipedia dump(2016) to find semantically related terms. We first average the embeddings of the words in the query to create a query vector. Next we search the embedding space for the 10 nearest terms that also occur in the top 10 documents.

The other 2 blackbox models we used were neural approaches:

\textbf{DESM}~\cite{mitra2016dual} models the relevance score of $d$ given $q$ as a parametrized sum between the syntactic relevance and semantic similarity, $P_{sem}$, between a learned query vector representation and the document vector representation. We select terms closest to the query vector that are also present in the top 50 documents of the initial result list as $\mathcal{G}_q$. To compute the vectors we employ the same glove embeddings from \textsc{EMB}. 

\textbf{DRMM} The Deep Relevance Matching Model~\cite{guo2016deep} utilizes pertained word embeddings to first create query-document term interaction count histograms. Additionally it includes a gating mechanism to learn which parts of the query to pay attention to. This is fed as input to a feed forward neural network that we trained with the Robust04 TREC adhoc retrieval test collection. We used glove embeddings (300 dimensions) trained on the same.

Note that since we cannot find $\mathcal{G}_q$ for DRMM, we examine it qualitatively instead in Section~\ref{sec:qualitative}

\mpara{Explanation engine}: While accuracy is measured against $\mathcal{G}_q$, fidelity however is sensitive to the choice of \explainer used to estimate $\mathcal{R}$. For the explanation engine we fix \explainer as a language model estimated using MLE with additive smoothing, i.e., $P(q_i|d) = \frac{count(q_i, d)}{|d|}$. We had to be careful in choosing a language model that is sufficiently different from the ranking function in $\mathcal{R}$ since we are only concerned with bag of words based models. 

\mpara{Other details.} We select the top 1000 terms as candidates for the strongly agnostic scenario. For the weakly agnostic case we use perturbation to filter out potential false positives. First we select the top 500 terms after reductive perturbation on a sample of documents. We further halve the candidate set using additive perturbation on the top-k documents-to-explain resulting in a final candidate set of 250 terms. Table~\ref{tab:rm3-10-weak} shows the recall of the candidate set before and after perturbation.

\section{Results}
\label{sec:experiments}

In this section we show both quantitative and qualitative results that demonstrate the efficacy of using query expansions and a simple ranker as explanations. The results and ensuing discussion are divided into 3 subsections. The first (Section~\ref{sec:sampling_results}) covers the results for the effect of sampling preference pairs from outside of the top-k results to construct the preference matrix. Section~\ref{sec:perturb_results} then describes the effect of perturbing documents in the weakly agnostic scenario. Finally in Section~\ref{sec:qualitative} we report anecdotal results for neural ranking models (DRMM and DESM) to demonstrate how our explanations can shed light on the intent and inherent training biases when using hard-to-interpret rankers.

\begin{table}
\footnotesize
\caption{\textsc{RM3-10} in a Strongly-Agnostic Setting (2500 features)} \label{tab:rm3-10} 

\begin{tabular}{lccc}
sampling & Accuracy & local fidelity & global fidelity \\

\midrule
\random & 0.3318 & 0.7011 & 0.5105 \\
\topkrandom & \textbf{0.5777} & 0.5000 & \textbf{0.7804} \\
\topkrankrandom & 0.5564 & 0.5274 & 0.7700 \\
\rankbiased & 0.3554 & 0.6427 & 0.5739 \\
\midrule
\topk & 0.1347 & \textbf{0.9576} & 0.3216 \\

\end{tabular}
\end{table}

\begin{table}
\footnotesize
\caption{\textsc{EMB} in a Strongly-Agnostic Setting (2500 features)} \label{tab:emb-prf} 

\begin{tabular}{lccc}
sampling & Accuracy & local fidelity & global fidelity \\

\midrule
\random & 0.1901 & 0.7405 & 0.4313 \\
\topkrandom & 0.7366 & 0.4973 & \textbf{0.8389} \\
\topkrankrandom & \textbf{0.7503} & 0.5508 & 0.8306 \\
\rankbiased & 0.2334 & 0.8070 & 0.4773 \\
\midrule
\topk & 0.0010 & 0.3109 & 0.1715 \\ 

\end{tabular}
\end{table}



\subsection{Effect of Sampling}
\label{sec:sampling_results}

From Table~\ref{tab:rm3-10}, we first observe the accuracy and local fidelity of \topk for \textsc{RM3-10}. This baseline constructs the preference matrix only using concordant pairs from the top-k results to explain. Local fidelity is 0.9576 indicating near perfect reproduction of the ranking produced by \bbranker. This result however stems from the presence of false positives terms in $T_q$ as indicated by the low accuracy value (0.1347). This is inspite of the tf-idf candidate selection that achieves a recall of 87\% (Table~\ref{tab:rm3-10-weak}). By selecting terms that maximize coverage, the greedy selection tends to overfit. For the query \textit{atypical squamous cells}, the $\mathcal{G}_q$ expansion terms for \textsc{RM3-10} are: atypical, tumor, lesion, cell, genotype, spindle and fibroblast. $T_q$ in this case however is: medicine, fibroblast, positive, cytoplasm, clear, different and damage. For the same query, when we use 2500 preference pairs instead of the 45 (45 pairs from top 10 documents) used by \topk, we see a steep increase in accuracy. $T_q$ when using \topkrandom is: atypical, cell, spindle, genotype, diethylstilbestrol and medicine. While these expansions are more accurate, we see a dip in local fidelity. Overall \topkrandom and \topkrankrandom achieve high accuracy at lower local fidelity. We also notice that by maximizing coverage of more pairs we increase global fidelity and as a result improve accuracy. We notice the same trend for \textsc{EMB} (Table~\ref{tab:emb-prf}) even though it uses an external source for expansions.

Carefully selecting pairs is also important for better explanations. Randomly selecting 2500 pairs (\random) leads to good accuracy with high local fidelity but \topkrandom and \topkrankrandom substantially improve accuracy (and in turn global fidelity). Both of these techniques allow for documents towards the bottom of the ranking to be selected. The benefit of doing so is observed when comparing against \rankbiased. The \rankbiased sampling achieves higher local fidelity than \topkrandom and \topkrankrandom but accuracy only slightly higher than \random. This indicates that sampling pairs from the top and bottom of the ranked list gives us the best "regularization" effect. We find that these trends hold for \textsc{RM3-20} and \textsc{EMB}. For \textsc{EMB}, \topkrankrandom achieves an accuracy of 0.75 and a local fidelity of 0.55 even though the expansion terms are selected from an embedding space learned on a different corpus. For queries with high accuracy, the loss in fidelity can be attributed to the limited capacity of \explainer. 

\begin{figure}[ht]
\centering
	\includegraphics[width=0.5\textwidth]{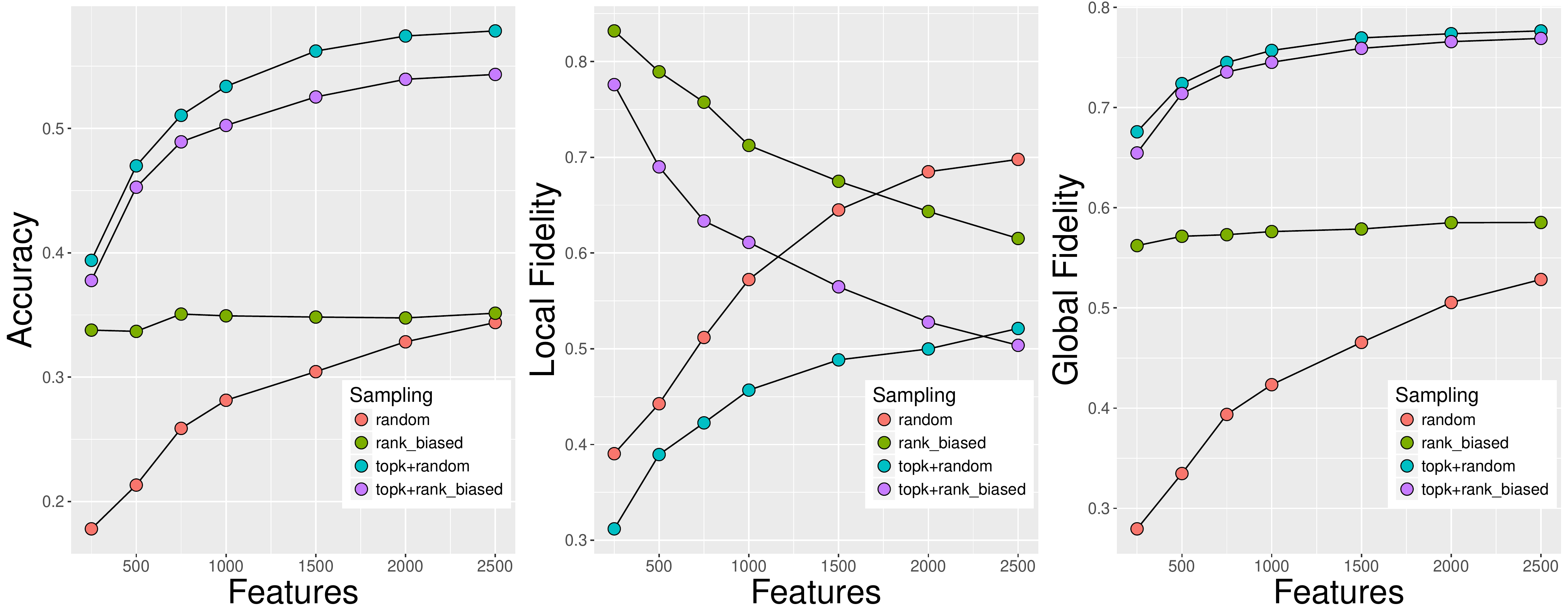}
    \caption{Effect of feature size when interpreting RM3-20 in strongly-agnostic setting. }.
    \label{fig:prf_tail}
\end{figure}

Creating preference matrices with many pairs can be computationally expensive (in the strongly agnostic scenario, these matrices are 1000 x 2500). Figure~\ref{fig:prf_tail} shows how accuracy varies with the number of sampled pairs for \textsc{PRF-20}. \rankbiased achieves its' highest accuracy with the fewest pairs. Sampling more pairs leads to no further increase. In general, adding pairs makes it more likely to have further evidence distinguishing true positives from false positives. In particular, utilizing fewer unique documents leads to relatively low accuracy that does not improve. \random on the other hand improves gradually which we also see in \topkrandom and \topkrankrandom. For \textsc{RM3-20}, when using \topkrandom and \topkrankrandom, we notice that adding more features has diminishing returns indicating 1500-2000 pairs to be optimal. Fidelity varies differently with the increasing pairs. For \random and \topkrandom, local fidelity gradually increases whereas it decreases for \rankbiased and \topkrankrandom. Adding random pairs helps better identify terms that can explain a preference pair. Adding more pairs from towards the top of the list leads to a preference matrix that is heavily skewed to the top pairs. Naturally the explanation from \rankbiased has high local fidelity.

\begin{table}
\footnotesize
\caption{\textsc{RM3-10}, \textsc{RM3-20}, \textsc{EMB} and \textsc{DESM} (500 features \& \topkrankrandom) in a Weakly-Agnostic Setting. Recall $C_I$ denotes recall of the initial 1000 candidate terms selected with Tf-Idf. Recall $C_{II}$ is the recall post perturbation based filtering.}
\label{tab:rm3-10-weak} 
\begin{tabular}{lccccc}
Model & Accuracy & local fidelity & global fidelity & recall $C_I$ & recall $C_{II}$\\

\midrule
\textsc{RM3-10} & 0.87 & 0.5777 & 0.7937 & 0.87 & 0.87 \\
\textsc{RM3-20} & 0.86 & 0.5685 & 0.8663 & 0.86 & 0.86\\
\textsc{EMB} & 0.86 & 0.6309 & 0.9263 & 0.86 & 0.86\\
\textsc{DESM} & 0.72 & 0.1862 & 0.4241 & 0.84 & 0.72\\

\midrule

\end{tabular}

\end{table}

\begin{table}
\footnotesize
\caption{Interpreting \textsc{DESM} in a Strongly-Agnostic and Weakly-Agnostic setting (500 features)} \label{tab:desm_0.9_noperturb} 

\begin{tabular}{lcccccc}
 & \multicolumn{2}{c}{Accuracy} & \multicolumn{2}{c}{local fidelity} & \multicolumn{2}{c}{global fidelity} \\
sampling & strong & weak & strong & weak & strong & weak \\

\midrule
\random & 0.1698  & 0.3273 & 0.3625 & 0.2941 & 0.2045 & 0.2551 \\
\topkrandom & 0.2914 & 0.5352 & 0.2288 & 0.2108 & 0.3775 & 0.3638 \\
\topkrankrandom & \textbf{0.3213} & \textbf{0.5534} & 0.2841 & 0.1890 & \textbf{0.3840} & \textbf{0.3687} \\
\rankbiased & 0.2331 & 0.4566 & 0.3028 & 0.2161  & 0.2785 & 0.3299 \\
\midrule
\topk & 0.1391 & 0.4066 & \textbf{0.6093} & \textbf{0.4070} & 0.1570 & 0.2737

\end{tabular}

\end{table}

\subsection{Effect of Perturbation}
\label{sec:perturb_results}

False positive terms that co-occur with true positives are difficult to remove in the strongly agnostic scenario. In the weakly agnostic setting, we can utilize \bbranker to find and remove such false positives from the initial candidate set of terms. For the non-DESM \bbranker chosen for our experiments, we can remove all false positive terms this way since score changes will only occur due to the terms in $\mathcal{G_q}$. Table~\ref{tab:rm3-10-weak} shows the results of interpreting the blackboxes in a weakly agnostic setting. Notice that the accuracy is the same as recall. However the fidelity values are not considerably higher. This can be attributed to the choice of \explainer. To tackle such scenarios, one can construct multiple preference pair matrices corresponding to varied \explainer. We leave this for future work.

However for \textsc{DESM}, false positives are harder to determine due to $P_{sem}$. Perturbation helps reduce the number of false positives (not completely remove them) with no loss to recall in \textsc{DESM} but as $\gamma$ decreases it is harder to distinguish a true positive from a false positive. Nonetheless, our approach can locally explain \textsc{DESM} with an accuracy of 0.5534 when using \topkrankrandom sampling which is similar to \textsc{RM3-10} (0.5777). Overall, we once again find \topkrandom and \topkrankrandom achieve high accuracy but lower local fidelity.

The major benefit of sampling is seen when considering \topk sampling. The reduced candidate set allows for relatively high accuracy of 0.4066 as opposed to the best value of 0.5534 for \textsc{DESM}. Comparing this to the strongly agnostic setting for the same, accuracy is 0.1391, nearly 3 orders of magnitude lower (Table~\ref{tab:desm_0.9_noperturb}). Here we also find that fidelity is similar in both agnostic settings further confirming that without perturbation overfitting increases. 

For the same query,\textit{atypical squamous cells}, the \textsc{DESM} $\mathcal{G}_q$ terms are: atypical, tumor, lesion, cell, tissue, brain and fibroblast. The predicted $T_q$ using \topkrandom is: brain, fibroblast, skull, cell, tissue, cellular and carcinoma. These expansions are more accurate and void of seemingly irrelevant terms like clear and different.




\subsection{Explaining DRMM}
\label{sec:qualitative}
\todo{
	maybe add the fidelity values too
}

\definecolor{lightgray}{gray}{0.9}
\begin{table}[ht!]
\caption{Anecdotal explanations} \label{tab:case_study} 
\footnotesize
\begin{tabular}{lr}
Query & Intent Explanation  \\
\midrule
\textbf{alexian brothers hospital} & \texttt{patient course war person} \\
(DRMM)&\texttt{sister leader alliance} \\
\textbf{alexian brothers hospital}& \texttt{medication treating memory } \\
(DESM)& \texttt{nurses father physical doctors}\\
\hline
\textbf{afghanistan flag} &  \texttt{US official inscription time} \\
(DRMM)&\texttt{transit dave november} \\
\textbf{afghanistan flag} &  \texttt{ symbol nation flagpole hoist} \\
(DESM)&  \texttt{ general banner flagstaff}\\
\hline
\textbf{fidel castro} &  \texttt{havana domestic cuba invest} \\
(DRMM)&\texttt{intestine real medical} \\
\textbf{fidel castro} &  \texttt{cuban havana dictator communist} \\
(DESM)&  \texttt{president raul gonzalez}\\
\hline
\textbf{how to find the mean} &  \texttt{x statistics plus know} \\
(DRMM)&\\
\textbf{how to find the mean}&  \texttt{actually say want meant} \\
(DESM)&\\
\hline
\textbf{battles of civil war} &  \texttt{official limit tennesse} \\
(DRMM)& \texttt{conflict army combat fought} \\
\textbf{battles of civil war}&\texttt{contain medic history} \\
(DESM)&  \texttt{iraq war end}\\
\hline
\textbf{electoral college 2008 results} &  \texttt{president popular statistic }\\
(DRMM)&\texttt{senate nominee gore}\\
\textbf{electoral college 2008 results} &  \texttt{election outcome expected} \\
(DESM)&  \texttt{2009 2004 following}\\

\end{tabular}
\end{table}
The goal of this case study is to demonstrate the utility of our explanation framework when interpreting complex models like DRMM and DESM (described in Section~\ref{sec:setup}) in a posthoc model agnostic setting. 
 

\textbf{Training Data Bias}. Table~\ref{tab:case_study} highlights queries that illustrate the power of our explanations. \textsc{DESM} uses wikipedia embeddings which is reflected in the more generic intent explanation terms (nurses as opposed to war for \textit{alexian brothers hospital}). Since DRMM was trained on Robust04, which is a news collection from 2004, we find more terms related to news-worthy events. This is particularly evident for \textit{afghan flag}. USA was involved in many conflicts in Afghanistan in the early 2000s and is promptly the reason why documents related to the USA get ranked higher for DRMM. \textsc{DESM} on the other hand favors documents more directly related to the concept of a flag. 

We also find evidence for temporal bias in the queries \textit{fidel castro} and \textit{electoral college 2008 results}.  DRMM ranks documents related to events in 2004 higher. Vice President and brother of Fidel, Raul Castro was handed control in 2006 (evidenced in \textsc{DESM}) due to Fidel Castro's illness which was a more prominent topic in 2004. Similarly DRMM considers documents related to Al Gore more relevant as compared to \textsc{DESM} for \textit{electoral college 2008 results}. 

\textbf{Model Bias}. The explanation also gives us insight into the nature of the ranker. For the query \textit{how to find mean}, even though the semantics of the query terms is resilient to temporal shifts, DRMM's gating mechanism helps capture the right semantics of the query.  \textsc{DESM}  on the other hand computes semantic similarity in a more simplistic manner, relying heavily on the pertained word embeddings to capture the right semantics.

\begin{figure}[ht]
\centering
	\includegraphics[width=0.33\textwidth]{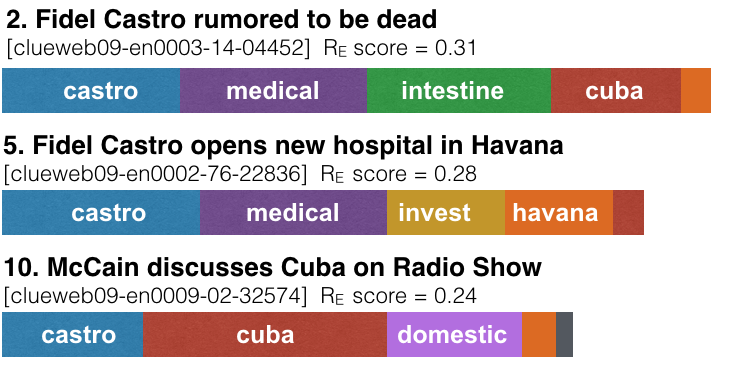}
    \caption{Explanation ($E_q$) for $d_2 > d_5 > d_{10}$ for $\mathcal{R}_{DRMM}$(\textit{fidel castro}). The length of a cell in the bar indicates term importance to $d$ as estimated by $E_q$ using \expscore{t}{d} which it estimates as a simple linear scoring function based on term frequency and document length here}
    \label{fig:castro-drmm}
\end{figure}

\textbf{Explaining pairs} The intent explanation terms when used with \explainer can further help us understand why a document was considered more relevant than another. Note that $E_q$ can only explain document pairs that are concordant in both target and explanation ranking. Figure~\ref{fig:castro-drmm} shows an anecdotal document pair explanation for the query \textit{fidel castro} and DRMM. Due to our choice of an \explainer that scores terms independently we can construct an easy-to-understand visual explanation that is a composition of term scores. 
In this ranking $d_2$ and $d_5$ seem to be similar when just looking at the explanation terms -- \texttt{havana domestic cuba invest intestine medical real} (both are related to medical issues). However on closer inspection, using \explainer, it becomes clear that \texttt{intestine} is a key term that is more prominent in $d_2$ than in $d_5$. Similarly, $d_{10}$ is ranked considerably lower since it only matches a few intent terms. 

\textbf{Summary:} Qualitatively we have seen how our approach can be used to identify temporal and model biases. Additionally we show how a visual aid can help explain pairwise rank differences. From the quantitative results we gathered that (i) using preference pairs only from the top-k results leads to high local fidelity but low accuracy. Sampling additional pairs from lower in the ranked list on the other hand can substantially increase accuracy since it indirectly optimizes for global fidelity at the cost of local fidelity. (ii) \topkrankrandom is usually the best sampling method which shows that taking pairs of documents that are mostly from the top with with finer differences (\rankbiased) or just randomly selected (\random) is not the best strategy. (iii) Document perturbation is key in the weakly agnostic setting to reduce the number of false positive terms. 




\section{Conclusion \& Future Work}

Complex retrieval models are effectively deployed as functional blacboxes in various domains. While they continue to be improved, little work has been to done on understanding and explaining the output of these ranking models. In this paper we detailed our framework for post-hoc explanations of black box rankers. In particular we focused on text-based rankers like DRMM and DESM which are hard to understand for developers and end users alike. Our proposed framework utilizes query expansions and a simple ranker to locally estimate the output of a bag-of-words based blackbox ranker. In the future we seek to interpret models that take external feature, sequence, proximity and position into account.


\newpage
\bibliographystyle{ACM-Reference-Format}
\bibliography{references} 

\end{document}